\input harvmac
\noblackbox

\input epsf

\newcount\figno
\figno=0
\def\fig#1#2#3{
\par\begingroup\parindent=0pt\leftskip=1cm\rightskip=1cm\parindent=0pt
\baselineskip=11pt
\global\advance\figno by 1
\midinsert
\epsfxsize=#3
\centerline{\epsfbox{#2}}
\vskip 12pt
{\bf Fig.\ \the\figno: } #1\par
\endinsert\endgroup\par
}
\def\figlabel#1{\xdef#1{\the\figno}}
\def\encadremath#1{\vbox{\hrule\hbox{\vrule\kern8pt\vbox{\kern8pt
\hbox{$\displaystyle #1$}\kern8pt}
\kern8pt\vrule}\hrule}}

\def\apm{{\alpha^{\prime}}}


\def\p{\partial}

\def\eqn#1#2{\xdef #1{(\secsym\the\meqno)}\writedef{#1\leftbracket#1}%
\global\advance\meqno by1$$#2\eqno#1\eqlabeL#1$$}


\lref\jmup{J. Maldacena, unpublished note.}
\lref\mpup{S. Minwalla and K. Pappododimas, unpublished.}

\lref\RecknagelIH{
A.~Recknagel and V.~Schomerus,
``Boundary deformation theory and moduli spaces of D-branes,''
Nucl.\ Phys.\ B {\bf 545}, 233 (1999)
[arXiv:hep-th/9811237].
}

\lref\mgas{
M.~Gutperle and A.~Strominger,
``Spacelike branes,''
JHEP {\bf 0204}, 018 (2002)
[arXiv:hep-th/0202210].
}
\lref\CallanUB{
C.~G.~Callan, I.~R.~Klebanov, A.~W.~Ludwig and J.~M.~Maldacena,
``Exact solution of a boundary conformal field theory,''
Nucl.\ Phys.\ B {\bf 422}, 417 (1994)
[arXiv:hep-th/9402113].
}

\lref\sz{A. Zamolodchikov, private communication.}

\lref\SenNU{
A.~Sen,
``Rolling tachyon,''
JHEP {\bf 0204}, 048 (2002)
[arXiv:hep-th/0203211].
}

\lref\aes{ A.~Strominger, ``Open string creation by s-branes,''
arXiv:hep-th/0209090.
}
\lref\TaniiUB{
Y.~Tanii and S.~I.~Yamaguchi,
``Two-dimensional quantum gravity on a disk,''
Mod.\ Phys.\ Lett.\ A {\bf 7}, 521 (1992)
[arXiv:hep-th/9110068].
}

\lref\SenIN{
A.~Sen,
``Tachyon matter,''
JHEP {\bf 0207}, 065 (2002)
[arXiv:hep-th/0203265].
}

\lref\GaberdielXM{
M.~R.~Gaberdiel, A.~Recknagel and G.~M.~Watts,
``The conformal boundary states for SU(2) at level 1,''
Nucl.\ Phys.\ B {\bf 626}, 344 (2002)
[arXiv:hep-th/0108102].
}

\lref\FateevIK{
V.~Fateev, A.~B.~Zamolodchikov and A.~B.~Zamolodchikov,
``Boundary Liouville field theory. I: Boundary state and boundary
two-point function,''
arXiv:hep-th/0001012.
}

\lref\TeschnerMD{
J.~Teschner,
``Remarks on Liouville theory with boundary,''
arXiv:hep-th/0009138.
}

\lref\JimboSS{
M.~Jimbo and T.~Miwa,
``QKZ equation with $|$q$|$=1 and correlation functions of the XXZ
model in the gapless regime,''
J.\ Phys.\ A {\bf 29}, 2923 (1996)
[arXiv:hep-th/9601135].
}

\lref\Barnes{E.W~Barnes, ``Theory of the Double Gamma function'',
Phil. Trans. Roy.Soc, {\bf A196} (1901) 265.}

\lref\GoulianQR{
M.~Goulian and M.~Li,
``Correlation Functions In Liouville Theory,''
Phys.\ Rev.\ Lett.\  {\bf 66}, 2051 (1991).
}

\lref\DornXN{
H.~Dorn and H.~J.~Otto,
``Two and three point functions in Liouville theory,''
Nucl.\ Phys.\ B {\bf 429}, 375 (1994)
[arXiv:hep-th/9403141].
}

\lref\ZamolodchikovAA{
A.~B.~Zamolodchikov and A.~B.~Zamolodchikov,
``Structure constants and conformal bootstrap in Liouville field theory,''
Nucl.\ Phys.\ B {\bf 477}, 577 (1996)
[arXiv:hep-th/9506136].
}
\lref\Okuda{T.~Okuda and S.~Sugimoto,
``Coupling of rolling tachyon to closed strings,''
Nucl.\ Phys.\ B {\bf 647}, 101 (2002)
[arXiv:hep-th/0208196].
} 
\lref\larsen{
F.~Larsen, A.~Naqvi and S.~Terashima,
``Rolling tachyons and decaying branes,''
arXiv:hep-th/0212248.
}

 \lref\hane{N. Hatano and D. Nelson, ``Vortex
Pinning and Non-Hermitian Quantum Mechanics,'',  Phys.\  Rev.\  B
{\bf 56}, 8651 (1997), [arXiv:cond-mat/9705290].}

\lref\afne{I. Affleck and D. Nelson, in progress.}

\lref\msx{A. Maloney, A. Strominger and X. Yin, in progress.}

\lref\PonsotNG{
B.~Ponsot and J.~Teschner,
``Boundary Liouville field theory: Boundary three point function,''
Nucl.\ Phys.\ B {\bf 622}, 309 (2002)
[arXiv:hep-th/0110244].}

\lref\FateevNN{
V.~Fateev, S.~Lukyanov, A.~B.~Zamolodchikov and A.~B.~Zamolodchikov,
``Expectation values of boundary fields in the boundary sine-Gordon  model,''
Phys.\ Lett.\ B {\bf 406}, 83 (1997)
[arXiv:hep-th/9702190].
}

\lref\RunkelNG{
I.~Runkel and G.~M.~Watts,
``A non-rational CFT with c = 1 as a limit of minimal models,''
JHEP {\bf 0109}, 006 (2001)
[arXiv:hep-th/0107118].
}

\lref\PolchinskiMY{
J.~Polchinski and L.~Thorlacius,
``Free Fermion Representation Of A Boundary Conformal Field Theory,''
Phys.\ Rev.\ D {\bf 50}, 622 (1994)
[arXiv:hep-th/9404008].
}

\lref\GreenGA{
M.~B.~Green and M.~Gutperle,
``Symmetry Breaking at enhanced Symmetry Points,''
Nucl.\ Phys.\ B {\bf 460}, 77 (1996)
[arXiv:hep-th/9509171].
}

\lref\FukudaBV{
T.~Fukuda and K.~Hosomichi,
``Super Liouville theory with boundary,''
Nucl.\ Phys.\ B {\bf 635}, 215 (2002)
[arXiv:hep-th/0202032].
}

\lref\Oku{
T.~Okuda and S.~Sugimoto,
``Coupling of rolling tachyon to closed strings,''
Nucl.\ Phys.\ B {\bf 647}, 101 (2002)
[arXiv:hep-th/0208196].
}

\lref\HosomichiXC{
K.~Hosomichi,
``Bulk-boundary propagator in Liouville theory on a disc,''
JHEP {\bf 0111}, 044 (2001)
[arXiv:hep-th/0108093].
}

\lref\GaberdielZQ{
M.~R.~Gaberdiel and A.~Recknagel,
``Conformal boundary states for free bosons and fermions,''
JHEP {\bf 0111}, 016 (2001)
[arXiv:hep-th/0108238].
}

\lref\ChenFP{
B.~Chen, M.~Li and F.~L.~Lin,
``Gravitational radiation of rolling tachyon,''
JHEP {\bf 0211}, 050 (2002)
[arXiv:hep-th/0209222].
}

\def\p{\partial}
\def\o{{\omega }}
\def\apm{{\alpha^\prime}}

\smallskip
\Title{\vbox{\baselineskip11pt
\hbox{hep-th/0301038}\hbox{SU-ITP-02/50}}}
{\vbox{\centerline{\bf{ Timelike Boundary Liouville Theory  }}
\vskip2pt }}

\centerline{Michael Gutperle\footnote{$^*$}{ Dept. of Physics,
Stanford University, Stanford CA. On leave of absence from Dept.
of Physics and Astronomy, UCLA, Los Angeles, CA.} and Andrew
Strominger\footnote{$^\dagger$}{Jefferson Physical Laboratory,
Harvard University, Cambridge MA }}

\bigskip

\baselineskip14pt

\centerline{\bf Abstract} {The timelike boundary Liouville (TBL)
conformal field theory consisting of  a negative norm boson with
an exponential boundary interaction is considered. TBL and its
close cousin, a positive norm boson with a non-hermitian boundary
interaction, arise in the description of the $c=1$ accumulation
point of $c<1$ minimal models, as the worldsheet description of
open string tachyon condensation in string theory and in scaling
limits of superconductors with line defects. Bulk correlators are
shown to be exactly soluble. In contrast, due to OPE singularities
near the boundary interaction, the computation of boundary
correlators is a challenging problem which we address but do not
fully solve. Analytic continuation from the known correlators of
spatial boundary Liouville to TBL encounters an infinite
accumulation of poles and zeros. A particular contour prescription
is proposed which cancels the poles against the zeros in the
boundary correlator $d(\o) $ of two operators of weight $\o^2$ and
yields a finite result.  A general relation is proposed between
two-point CFT correlators and stringy Bogolubov coefficients,
according to which the magnitude of $d(\o)$ determines the rate of
open string pair creation during tachyon condensation. The rate so
obtained agrees at large $\o$ with a minisuperspace analysis of
previous work. It is suggested that the mathematical ambiguity
arising in the prescription for analytic continuation of the
correlators corresponds to the physical ambiguity in the choice of
open string modes and vacua in a time dependent background.}
\smallskip

\Date{}
\listtoc
\writetoc

\baselineskip=20pt
\newsec{Introduction}

In this paper we study the two-dimensional conformal field theory
described by a $c=1$ negative norm
boson with an exponential interaction on the boundary. The action
is
\eqn\hlfba{S_{TBL}=-{1\over 2 \pi} \int_{\Sigma}
\partial X \bar
\partial X
  +{\lambda\over 2} \int_{\partial \Sigma}   e^{X}. }
We will refer to this as the TBL (timelike boundary Liouville)
theory. Due to the 'wrong' sign in front of the kinetic term in
\hlfba, the $X$ correlator on the upper half plane is\foot{Here and in
the rest of the paper we have set $\alpha'=1$.}
\eqn\xope{\langle X(z,\bar z)X(w, \bar w)\rangle  =  \ln |z-w|+
\ln|z-\bar w|,} and the boundary interaction is marginal. This
wrong sign also implies that the functional integral \eqn\wsc{\int
{\cal D} Xe^{-S_{TBL}}} is not well-defined. In order to define
the theory, we will need to specify some kind of analytic
continuation procedure.

TBL has not been previously studied in much detail.\foot{ A
minisuperspace analysis appeared in \aes. Certain
bulk one-point functions were recently computed in
\larsen.}  However it is a close cousin of several
theories which have been well-studied. Analytically continuing $X \to
i\phi$ we obtain a free positive-norm boson with a non-hermitian
boundary interaction \refs{\GaberdielXM,\RunkelNG}
\eqn\favc{S_{NH}={1\over 2 \pi} \int_{\Sigma}
\partial \phi \bar
\partial \phi
  +{\lambda\over 2} \int_{\partial \Sigma}   e^{i\phi}. }
This can be viewed as 'half' the boundary Sine-Gordon theory
(with a marginal boundary interaction)
\refs{\CallanUB,\PolchinskiMY,\FateevNN}, which has a   $\lambda e^{i\phi}+
\bar \lambda e^{-i\phi}$ boundary interaction. The correlators of
TBL are hence formally related to those following from \favc.
However since \favc\ has a non-hermitian interaction, its
correlators are also not unambiguously defined.

Generalizing the coefficient of the exponent in \favc, and
improving the stress tensor so that the interaction remains
marginal, we obtain the ordinary spacelike boundary Liouville
(SBL) theory \eqn\fbvc{S_{SBL}={1\over 2 \pi} \int_{\Sigma}
\partial \phi \bar
\partial \phi
  +{\lambda\over 2} \int_{\partial \Sigma}   e^{b\phi}. }
This theory has been studied for generic real values of $b$. The
two-point boundary correlators have been explicitly obtained in
\refs{\FateevIK,\TeschnerMD}, the bulk-boundary correlators are in
\HosomichiXC, an integral form of the three-point boundary
correlators has been given  in \PonsotNG, and supersymmetric
correlators are in \FukudaBV. We shall see that the analytic
continuation $b\to i$ from $S_{SBL}$ \fbvc\ to $S_{NH}$ \favc\ and
$S_{TBL}$ \hlfba\ is highly non-trivial and encounters ambiguities
for boundary correlators. This difficulty has been encountered
previously while studying the role of $S_{NH}$ in the $c=1$
accumulation point of minimal models \refs{\RunkelNG , \sz }. In
this paper we will give a specific, physically-motivated (in the
context of string theory) proposal for continuing the two-point
boundary correlator to TBL. The prescription involves approaching
the TBL theory through $c < 1$ theories with a linear dilaton.

Yet another related theory is the timelike
boundary Sine-Gordon theory
\eqn\hlfba{S_{TBSG}=-{1\over 2 \pi} \int_{\Sigma}
\partial X \bar
\partial X
  +\int_{\partial \Sigma} (\lambda_-e^{-X}+  \lambda_+e^{X}). }
The boundary state for this theory was found  by analytic
continuation from the spacelike case in \refs{\SenNU,\SenIN}, in the
context of 
string theory, where some cases describe an s-brane \mgas. Aspects
of closed string emission were computed in
\refs{\Oku,\jmup,\mpup,\ChenFP}. We expect the boundary
correlators for this theory, which have not been computed,  to be
more intricate due to the extra interaction term.  A further
complication is that there is no ``free'' region at $X\to -\infty$
in which the (open string) spectrum can be easily understood.
However the extra interaction term could also simplify matters by
controlling divergences and leading to a hermitian action for
$X\to i \phi$. We will not consider this interesting theory
further in the present paper.\foot{A minisuperspace analysis will
appear in \msx.}

The theories described by the actions $S_{TBL}$  and $S_{NH}$ are
of interest in a number of contexts. In string theory $S_{TBL}$ is
the worldsheet action describing time-dependent  open string
tachyon condensation \refs{\SenNU,\aes, \mgas}.  This can
equivalently be viewed as unstable D-brane decay or the future
half of an s-brane.  This relation will be further discussed in
section 2. Non-hermitian boundary interactions of the general
variety $S_{NH}$ \favc\ are realized in a variety of condensed
matter systems \hane.  $S_{NH}$ itself arises in the infrared
limit of a 2d superconductor with a magnetic field and a line
defect which are not parallel \afne. The non-hermiticity
corresponds to a lack of reflection symmetry across the defect.
Finally \favc\ is related to the $c=1$ theory obtained as the
accumulation point of the $c<1$ unitary boundary minimal models,
all of which it in a sense contains \refs{\RunkelNG , \sz }. This
highlights the
non-trivial nature of this CFT.

As noted above, the TBL theory is not well-defined without some
kind of additional prescription. This prescription should be
dictated by, and may depend on,  the physical context in which the
theory arises. In the context of string theory, we shall argue
that the two-point correlator gives stringy Bogolubov coefficients
describing quantum open string creation during tachyon
condensation. The creation rate depends only on the magnitude of
the two-point function, which was computed in the minisuperspace
approximation to TBL in \aes. We find that a natural
 prescription for defining the TBL two-point correlators
by analytic continuation gives a result in agreement with the
minisuperspace approximation at high energies. This connection
further suggests that the mathematical ambiguity in the correlator
corresponds physically to the ambiguity in the choice of a vacuum
state and modes for open strings during the time-dependent process
of tachyon condensation.\foot{It would be interesting to
understand the physical origin/resolution of these ambiguities in
the superconductor context \afne, where to date largely bulk
quantities have been considered.}

We wish to stress that we regard this work as a preliminary step
in defining the TBL CFT.  We have not given a procedure for
defining the boundary three-point function (known only in integral
form for the spacelike case), or verified that our prescription
yields correlators obeying the properties of a CFT. Indeed since
the TBL theory is not unitary it is not clear what those
properties should be. Further, we feel there is some hidden
``meaning'' in the (accumulation of) singularities which we have not
fathomed.  We regard all of these as interesting problems for
future investigation. Since tachyon condensation is an allowed
process in string theory we believe that, despite the apparently
singular behavior of the TBL theory, a well-defined set of
correlators should exist.

An intriguing feature of our continuation prescription to TBL is
the following. As mentioned above, an intermediate step involves
$c < 1$ timelike linear dilaton theories, which are of interest in
their own right. The proposed prescription determines the norm of
the boundary correlator $|d(\o)|$ for all real values of the
dilaton. Interestingly, as detailed in section 4.2, the  phase
${\rm Im} [\ln d(\o)]$ is determined only for 'rational' values of
the dilaton, and does not have a smooth extension to real values.

This paper is organized as follows. In section 2.1 the
minisuperspace analysis of TBL in the context of string theory is
reviewed. While the validity of the minisuperspace analysis is not
a priori obvious, it gives us invaluable clues as to which
operators to consider and what kind of phenomena to expect. In
section 2.2 a general relation between CFT two-point functions and
stringy Bogolubov coefficients is proposed. Section 3 describes
the computation of correlators of bulk operators, and explicitly
works out the one and two-point functions as well as the boundary
state. In section 4.1 we review the crucial results of \FateevIK\
on the boundary two-point function for ordinary boundary Liouville
theory which has a spacelike boson. Finally in 4.2 we detail our
proposal for continuing this two-point function to TBL. This
involves contours for analytic continuation of the background
charge (i.e. a timelike linear dilaton) and $\o$, as well as a
prescription to sum a certain series of residues after the
analytic continuation. The final result for the magnitude gives
agreement at high energies with the minisuperspace computation of
the open string creation rate. Properties of some special
functions appearing in the expressions for correlators are given
in an appendix.
\newsec{TBL and String Theory}

  TBL is the worldsheet description of a time dependent process
in which the open string tachyon field ${\cal T}=e^{X^0}$ starts
at its unstable minimum in the infinite past\foot{In classical
string theory the tachyon can be perched indefinitely at its
unstable minimum without being pushed off by quantum
fluctuations.} $X^0=-\infty$ and then rolls to an infinite value
in the far future $X^0=+\infty$. Such processes have been
discussed in \refs{\mgas, \SenNU, \aes}.  This may equivalently be
described as the decay of an unstable brane or the future half of
an s-brane (which consists of creation of an unstable brane
followed by its decay).

\subsec{Minisuperspace Review}
  The minisuperspace approximation to TBL was described in \aes\ and will be
reviewed in this subsection. While the validity of this
approximation is not a priori obvious, it nevertheless provides
invaluable clues as to what to look for in the exact treatment.

The $L_0=0$ constraint on the open string worldsheet for a half
s-brane becomes a Schr\"odinger equation for the open string wave
functions \aes\foot{In our conventions $\alpha^\prime=1$.}
\eqn\seql{\bigl( {\p^2 \over \p X^{2}} +{\lambda e^{X}}+{N-1
}+\vec p^2 \bigr)\psi(X)=0.} Here we abbreviate the timelike
coordinate $X^0$ as $X$, $\vec p$ is the spatial momentum and  $N$
is  the oscillator level number. The solutions to this are Bessel
functions \eqn\iio{\psi^{in}_{\vec p} ={ \lambda^{i\o} \over
\sqrt{ 2\o}}\Gamma(1-2i\o ) e^{i\vec p \cdot \vec x} J_{-2i
\o}(2\sqrt{\lambda}e^{X/2}),~~~~\omega \equiv \sqrt{N-1+\vec p^2}}
and their complex conjugates. In the far past this solution
approaches a positive frequency plane wave \eqn\jfh{X\to
-\infty,~~~~~~\psi_{\vec p}^{in}\to {1 \over \sqrt{2\o}}e^{-i\o
X+i\vec p \cdot \vec x} .} In the far future $X\to \infty$,
\eqn\zjfh{\psi_{\vec p}^{in}\to { \lambda^{i \o-1/4} \Gamma(1-2i\o
)\over \sqrt{8\pi  \o}} e^{- X/4+i\vec p \cdot \vec x}
\bigl(e^{{\pi \o}-2i\sqrt{\lambda}e^{X/2}+i {\pi \over
 4}}
+e^{{-\pi \o}+2i\sqrt{\lambda}e^{X/2}-i {\pi \over
 4}}  \bigr) .}
We see that the incoming modes $\psi_{\vec p}^{in}$ contain both
negative and positive frequency parts in the far future. This
indicates open string pair creation. Normalized outgoing positive
frequency modes are Hankel functions \eqn\otm{\psi_{\vec p}^{out}
= \sqrt{\pi \over 2 i}e^{-{\pi \o } +i\vec p \cdot \vec x}
H_{-2i\o}^{(2)}(2\sqrt{\lambda}e^{X/2}   ) \to {\lambda^{-1/4}
\over \sqrt{2}}e^{-{X \over 4}-2i\sqrt{\lambda}e^{X/2}+i\vec p
\cdot \vec x},~~~~X \to \infty.} The in and out modes are related
by the Bogolubov transformation \eqn\fda{\eqalign{\psi_{\vec
p}^{out}&=\alpha_{\vec p} \psi^{in}_{\vec p} +\beta_{\vec
p}\psi^{in*}_{-\vec p},\cr \psi_{\vec p}^{in}&=\alpha_{\vec p}^*
\psi^{out}_{\vec p} -\beta_{\vec p}\psi^{out*}_{-\vec p},\cr
 \alpha_{\vec p}&=  {\lambda^{-i\o}  \over\sqrt {4 \pi i \o}} \Gamma
 (1+2i\o)e^{\pi \o }, \cr
\beta_{\vec p} &=-{ \lambda^{i\o}\over\sqrt {4 \pi i \o}} \Gamma
(1-2i\o)e^{-{\pi\o } }.}} which obey $\alpha_{\vec
p}\alpha^*_{\vec p}-\beta_{\vec p}\beta^*_{\vec p}=1$ as required.
Expanding \eqn\expa{\phi=\sum_{\vec p} \bigl(\psi^{in}_{\vec p}
a^{in}_{\vec p} +\psi^{in*}_{\vec p }a_{\vec
p}^{in\dagger}\bigr)=\sum_{\vec p} \bigl(\psi^{out}_{\vec p}
a^{out}_{\vec p} +\psi^{out*}_{\vec p }a_{\vec
p}^{out\dagger}\bigr),} the in vacuum becomes
\eqn\ffd{|in>=\prod_{\vec p} (1-|\gamma^{in}_{\o}|^2)^{1/4}
e^{-\half\sum \gamma^{in}_{\o} (a_{\vec p}^{out\dagger})^2}|out>,}
where \eqn\wsx{\gamma^{in}_{\o}={\beta_{\vec p}^* \over
\alpha_{\vec p}}= -ie^{-2\pi \o},} where $\o$ and $\vec p$ are
related by \iio. The in vacuum is annihilated by $a_{\vec
p}^{in}=\alpha_{\vec p} a_{\vec p}^{out}+\beta^*_{\vec p} a_{-\vec
p}^{out\dagger}.$ Relation \ffd\ expresses the fact that if there
are no incoming particles at $X\to -\infty$, there will
necessarily be outgoing particles at $X \to \infty$.  $\alpha$ and
$\beta$ can be changed by phase redefinitions of the modes, but
the total string creation for a mode with frequency $\o$ is
characterized by $| \gamma_{\vec p} |$. Similarly the out vacuum
is an excited state of the in vacuum \eqn\ffo{|out>=\prod_{\vec p}
(1-|\gamma^{out}_{\o}|^2)^{1/4} e^{-\half\sum \gamma^{out}_{\o}
(a_{\vec p}^{in\dagger})^2}|in>,} where
\eqn\fsl{\gamma^{out}_\o=-{\beta_{\vec p}^* \over  \alpha^*_{\vec
p}} = {\lambda^{-2i\o} \Gamma(1+2i\o) \over \Gamma(1-2i\o)}
e^{-2\pi \o}.} The magnitude of this result will be reproduced for
large $\o$  in our CFT analysis of TBL.

\subsec{Two-Point Function as Stringy Bogolubov Coefficient}

  We would like to improve on the minisuperspace analysis
and obtain exact CFT results. Our first order of business is
to understand what correlator or other quantity in the CFT
determines the open string production rate.
   In order to understand this, we first review certain aspects
of the spacelike boundary Liouville (SBL) theory with action
\fbvc. Quantum states can be described as an incoming wave $e^{ip
\phi}$ from the free region $\phi\to -\infty$ which reflects off
the exponential $V \sim e^{b \phi}$ potential and returns as an
outgoing wave $d_b(p)e^{-ip \phi}$, where the reflection
coefficient $d_b(p)$ is a phase. The state then has the zero mode
dependence
 in the free region
\eqn\ffy{{\phi \to - \infty},~~~~~~\Psi_p(\phi) \to e^{-ip
\phi}+R_b(-ip)e^{ip \phi}.} Under the barrier, roughly speaking the
WKB wave function (for normalizable states) dies exponentially as
$exp ({-\sqrt V})\sim exp (-e^{b\phi/2})$, though of course the
theory is strongly coupled in this region so that statement is
heuristic. According to the operator state correspondence,
the reflection coefficient is given by
\eqn\rfc{R_b(-ip)=d_b({Q \over 2}-ip)}
where $Q=b+{1 \over b}$ and the two-point boundary
correlator on the upper half plane is \eqn\ff{d_b(\alpha )=<e^{\alpha
\phi}e^{\alpha
\phi}>_{SBL},} where the insertions are at $z=0$ and $z=1$.
This is illustrated in figure 1a.

\fig{(a) reflection amplitude for spacelike boundary Liouville
 and (b) analytically continued amplitude for TBL.}{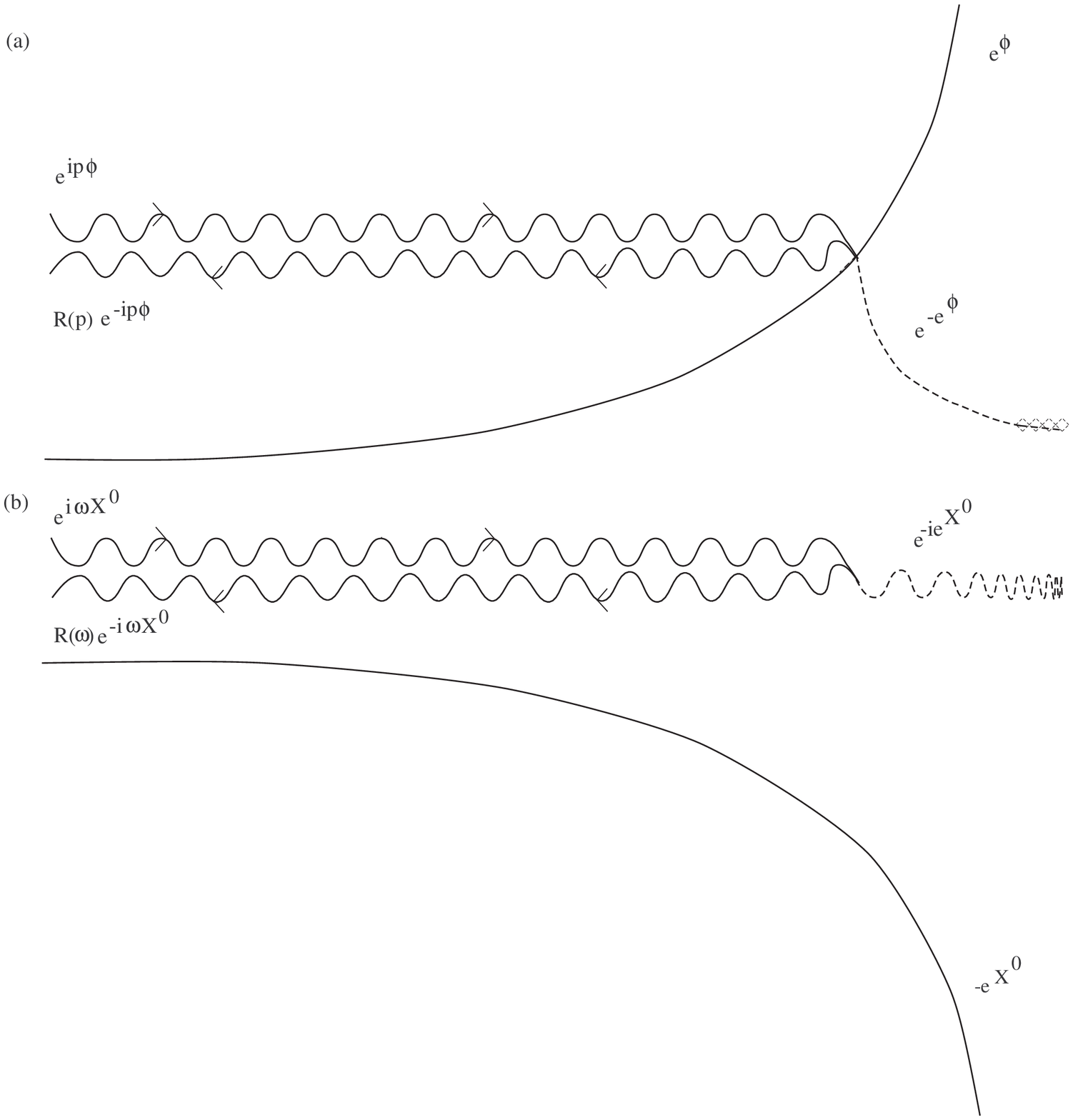}{5.0truein}

Let us now consider analytic continuation $\phi \to -iX$ from SBL
to TBL, so that $X$ is now a timelike target space coordinate, and
$p \to i\o$, as shown in figure 1b.
There is also an analytic continuation of the
screening charge such that $b\to i,~~~Q\to 0$ which will be detailed below.
Then in the free region $e^{ip \phi} \to e^{i\o X}$, and the wave
function behaves as\foot{Since $Q\to 0$ we need not distinguish between
$R_b$ and $d_b$ here.} \eqn\fzy{X \to - \infty,~~~~~\Psi_\o (  X) \to
e^{-i\o X}+d_i(\o)e^{i\o X},} where the appropriately continued
reflection coefficient \eqn\ssd{d_i(\o)=<e^{-i\o X}e^{-i\o X}>_{TBL}}
is no longer necessarily a pure phase.
 In the strong coupling region, the
potential is now negative relative to the kinetic term, and the
wave function behaves as  $exp ({-\sqrt V})\sim exp (\pm i
e^{X/2})$ (as indicated  in the minisuperspace result \zjfh ). In
the spacelike case, there are also two possible exponential
behaviors (growth and decay) but normalizability singles out the
decaying mode. Hence analytic continuation from SBL to TBL yields
a wave function with only one of the two asymptotic behaviors, as
opposed to a general admixture. Let us choose our prescription so
that this is the positive frequency outgoing wave. Then we
may interpret \fzy\ as the linear combination of incoming waves
that produces a purely positive frequency outgoing wave. Comparing
with \fda\ we then see that the two-point function is a ratio of
Bogolubov coefficients \eqn\frhj{d_i(\o)={\beta_\o \over
\alpha_\o}.} The string pair
production rate is determined by the magnitude of $d$. 
In the minisuperspace approximation
\eqn\dkaw{
|d_i(\o)|=e^{-2\pi \o}.}

There is another way of understanding the relation \frhj.
Mathematically, correlators in TBL are not unique because one must
specify an analytic continuation procedure. Physically they are
not unique because one must specify the vacuum state in a
time-dependent background. As discussed above, 
analytic continuation from SBL to TBL
most naturally gives correlators in the out vacuum. The out vacuum
is  represented as an excited state of the in vacuum in
expression \ffo. One then easily finds that, in this state,
 the minisuperspace S-matrix for scattering  two incoming strings to
 zero outgoing
strings is given by
\eqn\vvn{ -\gamma_\o^{out *}={\beta_\o \over
\alpha_\o}=d_i(\o).} Hence one may interpret the two
point function \ssd\ as giving this S-matrix element.

It is natural to conjecture that this relationship extends beyond
the example discussed here. More generally we expect that the disk or sphere
two-point function
for open or closed strings in a time dependent background gives the
stringy Bogolubov coefficients.

Now we turn to computation of the CFT correlators.

\newsec{Bulk Correlators}

The TBL theory is a  boundary deformation of a free timelike boson
on the upper half plane
\eqn\halfba{S_{TBL}=-{1\over 2 \pi} \int_{\Sigma}
\partial  X
\bar \partial X
  +{\lambda\over 2} \int_{\partial \Sigma}  e^{X} }
A correlator involving $n$ closed string vertices and $m$ open
string vertices is formally given by the path integral
\eqn\halbc{\eqalign{A&=\langle \prod_{i=1}^n e^{\beta_iX(z_i,\bar
z_i)} \prod_{j=1}^m e^{\gamma_j X(x_j)}\rangle\cr &= \int DX
e^{-S} \prod_{i=1}^n e^{\beta_iX(z_i,\bar z_i)} \prod_{j=1}^m
e^{\gamma_j X(x_j),}}} where $x_i$ is a point on the real axis.
Following \GoulianQR , we decompose $X=q+\hat X$ where $q$ is the
zero mode of $X$. Then the $q$ integral can be done exactly,
yielding \eqn\halfbb{A= \Gamma(-s) \left({\lambda\over 2}\right)^s
\langle \prod_{i=1}^n e^{\beta_i\hat X(z_i,\bar z_i)}
\prod_{j=1}^m e^{\gamma_j \hat X(x_j)} \Big( \int dy\; e^{\hat
X(y)}\Big)^s\rangle_{Free}, } where $s= -\sum_j^n \beta_j
-\sum_{i}^m \gamma_i$. For integer $s$, $\Gamma(-s)$ has a simple
pole and the residue is given by the integral over  the nonzero
modes, which can be evaluated using the free field correlation
function on the half plane with Neumann boundary conditions.
(A useful reference for the resulting integrals is \TaniiUB).The
general correlator is then obtained by analytic continuation in
$s$. However, since the residues can be perturbatively evaluated
only for integer $s$, and analytic continuation from the integers
in not unique, the final answer must be checked using various
consistency conditions such as factorization, crossing symmetry, etc.
   This technique was
used in the calculation of the three point function for the bulk
Liouville theory \DornXN\ZamolodchikovAA.

This procedure however is a bit problematic for open string
correlators because of singularities when the boundary operator
and interaction insertions coincide. A prescription must be
specified for dealing with these correlators. Ultimately we
believe that this corresponds to the ambiguity in the choice of
open string vacuum, to which closed string correlators on the disk
are insensitive.

Rather than directly computing the integrals in \halfbb, bulk
correlators of the form \halbc\ may alternately be evaluated using
contour deformation.  TBL has a level
one SU(2) current algebra generated by\foot{ In our conventions
$X(z,\bar z)=\half
(X(z)+X(\bar z))$,  $X(z)X(w)\sim 2\ln (z-w)$ and $\apm=1$.}
\eqn\rftg{j_\pm(z) =e^{\pm
X(z)},~~~~~~j_3(z)={1 \over 2}\p X(z) ,} which obey
\eqn\opes{\eqalign{j_-(z)j_+(w)&\sim {1 \over (z-w)^2}-{2j_3(w)
\over (z-w)},\cr j_3(z)j_\pm(w)&\sim \pm{j_\pm(w) \over
(z-w)},\cr}} Note however that in the
standard norm for a timelike boson $X$,
$j_3$ is anti-hermitian while $j^\pm$ are both
hermitian. Nevertheless the charges \eqn\cdgf{J_\pm=\oint {dz
\over 2\pi i} j_\pm(z), ~~~~~~J_3=\oint {dz \over 2\pi i} j_3(z),
} obey the usual commutation relations
\eqn\fcv{[J_-,J_+]=-2J_3,~~~~~~[J_3,J_\pm]=\pm J_\pm } and are
therefore useful for computing correlators. To exploit this we
note that the TBL boundary interaction is simply
\eqn\rdx{{\lambda\over 2} \int_{\partial \Sigma} d\tau  e^{X}
=i \pi \lambda  J^+.} Correlators may then be evaluated by e.g.
deforming the $J^+$ contour into the upper half plane and letting
it act on the operator insertions as in \CallanUB.\foot{Life is
not so simple with boundary insertions because one must specify
the contour prescription near the boundary operator insertion.}

\subsec{One-Point Function}
In this section we
calculate the one-point function of the closed string vertex operator $e^{-
n X}$
inserted at the center of a unit disk. Because of momentum conservation the
perturbative contribution is
given by the insertion of $n$ Liouville boundary interaction terms.
\eqn\onepta{\eqalign{\langle e^{- n X}(z,\bar z)\rangle_{TBL} &= \left({\lambda
\over 2}\right)^n \langle
e^{- n X}(z,\bar z)
\prod_{i=1}^n \int dx_i e^{X}(x_i)\rangle\cr
&= \left({\lambda\over 2}\right)^n  |z-
\bar z|^{n^2\over 2} \prod_{i=1}^n \int dx_i
\prod_{i<j}|x_i-x_j|^2 \prod_{i=1}^n |z-x_i|^{-n}|\bar z-x_i|^{-n}
\cr &= \left({\lambda\over 2}\right)^n   |z-
\bar z|^{-{n^2\over 2}} \int
\prod_{i=1}^s {du_i\over u_i} \prod_{i<j}^s |u_i-u_j|^2 \cr &=
 |z-
\bar z|^{-{n^2\over 2}} \left({\lambda\over 2}\right)^n (2\pi)^n
\Gamma(n+1).}}
Where is the third line the upper half plane was mapped to the disk, which
mapped the integrations to the well known Dyson-gas form. The perturbative
result \onepta\ can now be used to determine the general form of the bulk
one-point function by analytical continuation using \halfbb.
\eqn\analconh{\eqalign{\langle e^{\beta X}(z,\bar z)\rangle_{TBL}&= |z-
\bar z|^{-2h_\beta}(\pi \lambda)^{-\beta} \Gamma(
\beta)\Gamma(1-\beta)\cr
&=  |z-
\bar z|^{-2h_\beta}(\pi \lambda)^{-\beta} {\pi \over \sin \pi \beta}. }}
The one-point function \analconh\ can now be analytically continued
via $\beta\to -i\omega$, giving 
\eqn\analconb{\langle e^{-i\omega X}(z,\bar z)\rangle_{TBL} = |z-
\bar z|^{-2h_\omega}(\pi \lambda)^{i\omega} {\pi \over  i
}{1\over\sinh \pi \omega} .}

\subsec{Boundary States} The collection of all bulk one-point
functions can be represented by a boundary state. The boundary
state $\mid B\rangle_{BSG}$ for the boundary Sine-Gordon theory
\eqn\cklma{S={1\over 2 \pi} \int_{\Sigma} \partial \phi \bar
\partial \phi
  +{1\over 2} \int_{\partial \Sigma}  \big( \lambda e^{i}\phi+\bar
  \lambda e^{-i\phi}\big)}
was found using the bulk $SU(2)$ current algebra in \CallanUB (see
  also \refs{\PolchinskiMY , \GreenGA ,\RecknagelIH, \GaberdielZQ}). For a
noncompact boson one has (up to normalization)
\eqn\cklmb{ \mid B\rangle_{BSG}=
\sum_{j}\sum_{m=-j}^j D^j_{m,-m} \big(
  g(\lambda,\bar \lambda)\big) \mid j;m,m\rangle\rangle.}
Here $\mid j;m,m\rangle\rangle$ is the Ishibashi state associated with
the $SU(2)$ primary field $\mid j;m,m\rangle$.   $D^j_{m,-m}(g)$ is
the spin $j$ representation of the $SU(2)$ rotation given
by
\eqn\cklmc{g(\lambda,\bar \lambda) = e^{i\pi(\lambda J_++\bar
  \lambda J_-)}=\pmatrix{\cos(\pi |\lambda|)&
  i  \lambda {\sin (\pi|\lambda|)\over |\lambda|} \cr
 i\bar \lambda {\sin (\pi|\lambda|)\over |\lambda|}&\cos(\pi
 |\lambda|) }.}
Reality of the boundary interaction demands that $\lambda$ and
$\bar\lambda$ are complex conjugates. Sen \refs{\SenNU,\SenIN} observed that
an analytic
continuation $\phi \to -iX^0$ produces an exact time dependent open
string background.

It was pointed out in \GaberdielXM\ (and at intermediate stages of
the calculations in \CallanUB) that the boundary states can also
be constructed for $g\in SL(2,C)$, and in particular we can set
$\bar \lambda=0$ to obtain the non-hermitian theory
whose action $S_{NH}$ is in \favc.
The unitary rotation matrix \cklmc\  becomes
a raising operator. The boundary state becomes simply
\eqn\dndsta{\mid B\rangle_{NH} = \sum_j \sum_{m\geq 0}^j
  \pmatrix{j+m\cr 2m} (i\pi\lambda)^{2m} \mid j; m,m\rangle\rangle.}
Inspection of the $SU(2)$ currents \rftg\ of TBL theory reveals that
under $\phi \to -iX^0$, $J_k \to J_k$ and hence $\mid j;
m,m\rangle\rangle
\to \mid j; m,m\rangle\rangle$. Therefore we may also write
\eqn\dsta{\mid B\rangle_{TBL} = \sum_j \sum_{m\geq 0}^j
  \pmatrix{j+m\cr 2m} (i\pi\lambda)^{2m} \mid j; m,m\rangle\rangle.}

Following a related discussion in \SenNU, the component of the
boundary state \dsta\ which does not contain
any oscillator modes can be obtained by setting $m=j$:
\foot{As in \SenNU\ there are extra phases $j^j$ appearing
  in  $\mid j; j,j\rangle$ which can be fixed by demanding that the
  $\lambda=\bar\lambda=1/2$ state corresponds to an array of D0-branes.}
\eqn\dndstb{\eqalign{\mid B\rangle_0 &= \sum_{j} (i\pi\lambda)^{2j}
    \mid j; j,j\rangle \cr
&= \sum_{n=0}^\infty (-\pi \lambda)^n e^{nq}\mid 0\rangle\cr
&= {1\over 1+\lambda \pi e^{q}} \mid 0\rangle.}}
This result agrees with the appropriate limit of the more general boundary
state found in \SenNU.
Although every term in the second line of \dndstb\ diverges at late
times the resummed expression is well behaved, in particular there
exists a Fourier transform which gives the closed string one-point function
\eqn\dndstc{\langle e^{2i\omega X} \mid B\rangle= const (\pi
  \lambda)^{2 i\omega} {1\over \sinh (2 \pi \omega)},}
in agreement with \analconb.

\subsec{N-Point Correlators} Perturbative correlation functions
involving only bulk vertex operators can be easily evaluated using
contour deformation techniques. For example the two-point function
is \eqn\twptg{\eqalign{A_2(j_1,j_2)&=\langle e^{-2j_1
X}(z_1,\bar z_1) e^{-2j_2 X}(z_2,\bar z_2)\rangle_{TBL} \cr &=
(2\pi i)^{2(j_1+j_2)}\left({\lambda \over 2}\right)^{2(j_1+j_2)}
{1\over (2(j_1+j_2))!}\cr &\times\langle e^{-j_1 X}(z_1) e^{-j_1
X}(\bar z_1) e^{-j_2 X}(z_2) e^{-j_2 X}(\bar z_2)
\prod_{i=1}^{2(j_1+j_2)}\oint {dx_i\over 2\pi i} e^X(x_i)\rangle
\cr &=(2\pi i)^{2(j_1+j_2)}\left({\lambda \over
2}\right)^{2(j_1+j_2)}\langle e^{-j_1 X}(z_1) e^{+j_1 X}(\bar z_1)
e^{-j_2 X}(z_2) e^{+j_2 X}(\bar z_2)\rangle \cr &= (2\pi
i)^{2(j_1+j_2)}\left({\lambda \over 2}\right)^{2(j_1+j_2)}
|z_1-\bar z_1|^{-{j_1^2\over 2}}  |z_2-\bar z_2|^{-{j_2^2\over 2}}
|z_1-z_2|^{j_1j_2} |z_1-\bar z_2| ^{-j_1j_2}. }} In the
second line the bulk vertex operators on the half plane where
split into holomorphic and antiholomorphic parts on the plane
using the doubling trick. Then the  contours along the real axis
were pulled off the lower half plane and the $SU(2)$ algebra \fcv\
was used to turn $e^{-j_i X}(\bar z_i)$ into $e^{+ j_i X}(\bar
z_i)$. Note that all combinatorial factors cancel in the end. 
It is straightforward
to generalize the contour deformation techniques to evaluate bulk
N-point functions. Hence as far as the bulk correlation functions are
concerned the TBL theory is very simple. We shall
see that this is not the case for correlation functions involving
boundary vertex operators.

\newsec{Boundary Correlators}

As mentioned above, the simple methods for computing bulk
correlators encounter ambiguous singularities when applied to
boundary correlators. In this section we will define the two-point
correlator by analytic continuation from known expressions for the
two-point correlator of the spatial boundary Liouville theory
\refs{\FateevIK, \TeschnerMD}.  We shall see that even this
procedure is ambiguous: an infinite number of pairs of poles and
singularities accumulate at precisely the point we wish to
continue to.  We will adopt a simple (but not obviously unique)
prescription in which these poles and singularities cancel one
another and a finite answer is obtained for the two-point
correlator.

\subsec{Spacelike Boundary Liouville} Spacelike boundary Liouville
theory can be defined by the following action on the half plane:
\eqn\blouva{S_{SBL}= {1\over 2\pi}\int_{\Sigma} \big(\partial \phi
\bar
\partial \phi+ \pi \mu
e^{2b \phi}\big) + {\lambda \over 2}\int_{\partial \Sigma}
e^{b\phi}.} Here $\mu$ and $\lambda$ are the bulk and boundary
cosmological constants respectively. The Liouville coupling
constant $b$ determines the background charge $Q=b+{1\over b}$ and
the central charge $c=1+6Q^2$ of the theory. Boundary vertex
operators $e^{\o \phi}$ have conformal dimension $h_\o=\o(Q-\o)$.
There are two important quantities calculated by Fateev,
Zamolodchikov and Zamolodchikov  \FateevIK (see also \TeschnerMD).
Firstly the bulk one-point function
\eqn\oneptb{\eqalign{U(\alpha)&=(z-\bar z)^{2h_\alpha} \langle
e^{2\alpha X (z, \bar z)}\rangle\cr &= {2\over b} \big(\pi \mu
\gamma(b^2) \big)^{Q-2\alpha\over 2b}\Gamma(
2b\alpha-b^2)\Gamma({2\alpha\over b}-{1\over b^2}-1)\cosh\big((
Q-2\alpha)\pi s\big),}} where $\gamma(x)=\Gamma(x)/\Gamma(1-x)$.
Secondly the  boundary two-point function \eqn\blouvb{\langle
e^{\o_1 \phi (x)} e^{\o_2\phi (0) }\rangle= {1\over |x|^{2
h_\o}}\Big(
  \delta(Q-\o_1-\o_2) +
  \delta(\o_1-\o_2) d(\o)\Big),}
where \eqn\blvc{ d(\o)= \Big( \pi \mu \gamma(b^2)
  b^{2-2b^2}\Big)^{Q-2\o\over 2b} {G_b( Q-2\o)\over
  G_b(2\o-Q)} {1\over S_b(\o+is)S_b(\o-is)S_b(\o)^2}.}
Here $G_b$ and $S_b$ are special functions defined in \FateevIK\
and  related to the Barnes
  double Gamma function \Barnes\ (see Appendix A  for a collection
  of  useful formulae). The parameter $s$ is related to the
  coupling constants in \blouva\
  of the theory in the following way
\eqn\blouvd{\cosh^2(\pi b s)= {\lambda^2 \over 4 \mu}\sin (\pi b^2).}

Our current interest is the case for which the bulk cosmological
constant is turned off. From \blouvd\ it follows that as $\mu \to
0$ one has to take $s\to \infty$. \eqn\blouve{\lim_{s\to \infty}
U(\alpha)= {1\over b} \Big(  {\pi \lambda \over
\Gamma(1-b^2)}\Big)^{Q-2\alpha\over b}\Gamma(
2b\alpha-b^2)\Gamma({2\alpha\over b}-{1\over b^2}-1). } Using
\blouvd\ and A.11 the two-point function has the limit\foot{This formula
appears in \FateevNN, but apparently with a different power of 2 in the 
normalization.}
\eqn\blouvf{\lim_{s\to \infty} d(\o)\equiv d_b(\o)= \Big(
 {\pi \lambda b^{1-b^2}\over \Gamma(1-b^2)}\Big)^{Q-2\o\over b}
{G_b(Q-2\o)\over G_b( 2\o-Q)}{1 \over S_b(\o)^2}.} The SBL theory
with interaction $e^{b\phi}$ can (at least formally) be related to
the TBL theory with interaction $e^{X}$ by taking $b\to i$ while
$\phi \to -iX$. Note that in this limit $Q\to 0$, $c\to 1$ and one
gets a free boson with vanishing background charge. Furthermore
perturbative correlation functions are clearly identical for the
two theories.

For the bulk one-point function \blouve\ one finds
\eqn\blouvg{\lim_{b\to i}\lim_{s\to \infty} U(\alpha) = \pi ({\pi
\lambda })^{2i\alpha} {1\over  \sinh (2\pi \alpha)},} which (up to
normalization) agrees with \dndstc\
for $\alpha=\omega$.

\subsec{Analytic Continuation to Timelike Boundary Liouville}

We wish to obtain the TBL two-point function from the SBL two
point function \blvc\ by the analytic continuation \eqn\acnt{
\langle e^{-i\o X}e^{-i \o X}\rangle_{TBL}\equiv \langle e^{\o
\phi}e^{ \o \phi}\rangle_{SBL, b=i}=d_i(\o).} This however is not
as straightforward as it sounds. As seen in the appendix, the
special functions $G_b(z)$ and $S_b(z)$ appearing in \blvc\ have
poles and/or zeros at $z=mb+{n \over b}$ for integer $m$ and $n$.
If we take $b\to i $ from the real axis along the unit circle
these poles/zeros are at $z=(m+n){\rm Re} b+(m-n){\rm Im} b$, and
an infinite number of them accumulate at every integer multiple of
$i$.\foot{This singularity may be related to the accumulation of
boundary minimal models at $c=1$ \refs{\sz,\RunkelNG}.} For this reason
$G_b$ is not defined for $b=i$ \Barnes.

However it turns out that if we look at the particular ratio of
special functions appearing in $d_b(\o)$ \blouvf, we shall see
that the poles and singularities accumulate in pairs and can be
arranged to cancel for real $\o$. This will enable us to give a
prescription defining $d_b$. Using recursion and  product relations
from the appendix, the ratio of special functions appearing in
$d_b$ is \eqn\prode{\eqalign{{G_b(Q-2\o)\over G_b( 2\o-Q)}&{1
\over S_b(\o)^2}\cr &= {G_b(Q-2\o)\over G_b(2\o-Q)}
  {G_b(\o)^2\over G_b(Q-\o)^2}\cr
&=Y_b(\o)b^{ {2\o\over b}- 2b\o-{1\over
  b^2}+b^2 } { \Gamma({2\o\over b}-{1\over b^2}) \Gamma(2\o b -1
  -b^2)\Gamma(-{2\o\over b} +1) \Gamma(-2 \o b)\over \Gamma^2(
  -{\o\over b}+1) \Gamma^2(-b \o)}, }}
where
\eqn\yb{Y_b(\o)\equiv {G_b(-2\o)\over G_b(2\o)}
  {G_b(\o)^2\over G_b(-\o)^2}.}
 Using the product representation
A.10 one has simply \eqn\prodx{\eqalign{ Y(\o)&=
\prod_{m=0}^\infty
  \prod_{n=0}^{\infty }\big({2\o+\Omega \over -2\o
  +\Omega}\big)\big({-\o+\Omega \over \o +\Omega}\big)^2}.}
where $\Omega= m b +n/b$. It can be seen that
the product is absolutely convergent for generic complex $b$.

We now wish to understand the behavior of this correlator
for $b \to i$. We will take $b \to i$ by first going to the imaginary axis,
so that $b=i\beta$ and $Q=i(\beta-{1 \over \beta})$ with $\beta$ real,
and then taking $\beta \to 1$. Physically this corresponds to adding a
real timelike linear dilaton which alters the growth of the tachyon.
For pure imaginary $b$, $\Omega$ is also pure imaginary, and
(for real $\o$) the
factors in \prodx\ appear in complex conjugate pairs. Hence for this case 
$Y$ is formally a pure phase. In order to make a more precise statement
and determine the phase we now introduce
the integral form of the special functions.

In \FateevIK\ one finds the integral representation \eqn\woph{\ln
S_b(x)=\half\int_0^{\infty}{dt \over t}\bigl[{\sinh (Q-2x)t \over
\sinh(bt) \sinh (t/ b)}+{2x-Q \over t}\bigr].} We take ${\rm
Im}b>0,~{\rm Re}b>0$ and $0<2x<Q$ with $x$ real, which implies
convergence of \woph. Other values of the parameters will be
obtained by analytic continuation. By deforming the integration
contour, \woph\ may be rewritten \eqn\ftg{\ln S_b(x)= I_{b}(x)
+\Sigma_b(x).} as the sum of an integral $I_{b}(x)$ over the
positive imaginary axis plus a sum $\Sigma_b(x)$ of simple pole
residues at $t={n \pi i \over b}$.\foot{We might also have
deformed to the negative imaginary axis which would have picked up
the poles at $t=-n\pi i b$ and changed the results below by the
replacement $b\to {1 \over b}$. One possibility is to take half
the sum of the two contours which would manifestly preserve the
$b\to{1 \over b}$ symmetry. However there is no change in the
final  formulae for $b=i$, which is our main interest here, so we
will not further explore these alternate prescriptions.}${}^{,}$\foot{In
the appendix of \RunkelNG\ it was suggested that the residue sum
might be dropped in determining the correlators. That leads to 
correlators which are pure phases of constant magnitude. } (The
contribution from the quarter-circle at infinity vanishes.)
Defining $t=i\tau$ the integral is \eqn\intr{I_{b}(x)={i \over
2}\int_0^{\infty}{d \tau \over
\tau}\bigl[{\sin\big((2x-Q)\tau\big) \over \sin (b \tau) \sin
(\tau / b)}+{Q-2x \over \tau}\bigr].} For $b \to i\beta$ with
$\beta$ real and positive, this reduces to the convergent
expression \eqn\nxtr{I_{i\beta}(x)={i \over 2}\int_0^{\infty}{d
\tau \over \tau}\bigl[{\sin\big((2x-Q)\tau\big) \over \sinh (\beta
\tau) \sinh (\tau /\beta) }+{Q-2x \over \tau}\bigr].}  For $b \to
i$, this further reduces to \eqn\ntr{I_{i}(x)={i \over
2}\int_0^{\infty}{d \tau \over \tau}\bigl[{\sin(2x\tau) \over
\sinh^2(\tau) }-{2x \over \tau}\bigr],} which contributes a pure
phase to $S$. The sum over pole residues is
\eqn\svc{\eqalign{\Sigma_b(x) &=\sum_{n=1}^{\infty}{(-)^n\sin
\left({ \pi n (Q-2x) \over b}\right) \over n \sin \left({\pi n
\over b^2}\right)}\cr &=\sum_{n=1}^{\infty}{ 1 \over n} \bigl[
\cos\left( {2\pi nx \over b}\right) -\cot \left( {\pi n\over
b^2}\right)\sin\left( {2\pi n x \over b}\right) \bigr].\cr }} In
order to take $b \to i\beta$ in this expression, we define
\eqn\ssz{{1\over b^2}= -{1 \over \beta^2}+i \epsilon,~~~~~
x=-iby,} where $y$ is real.\foot{ For $b$ on the imaginary axis,
reality of $x$ and reality of $y$ are the same thing. Keeping $y$
rather than $x$ real for $b$ off the imaginary axis simplifies the
calculations.} The real part of the sum  is then
 \eqn\flq{{\rm Re} [\Sigma_{i\beta,\epsilon}]=\sum_{n=1}^{\infty}{ 1 \over n}\bigl[
\cosh (2 \pi n y) +\sinh (2\pi n y ){\rm Im}[\cot ({n\pi \over
\beta^2}-n\pi i\epsilon)] \bigr].} We wish to take $\epsilon \to
0$ with $y$ fixed, which takes us outside the radius of
convergence of \flq. The problematic terms for small $\epsilon$
are the ones that behave as $e^{2\pi ny}$ ($e^{-2\pi n y}$) for
positive (negative) $y$, ie. the first (second) term in the
expression \eqn\fzlq{{\rm Re}
[\Sigma_{i\beta,\epsilon}]=\sum_{n=1}^{\infty}{ e^{2\pi ny} \over
2n}\bigl[ 1+{\rm Im}[\cot ({n\pi \over \beta^2}-n\pi
i\epsilon)]\bigr]
 + \sum_{n=1}^{\infty}{ e^{-2\pi ny} \over
2n}\bigl[ 1-{\rm Im}[\cot ({n\pi \over \beta^2}-n\pi i\epsilon)]
\bigr].} When $\epsilon \to 0$, we  will define the first (second)
term for positive (negative) real $y$ by analytic continuation
from negative (positive) real $y$. The resulting expression is
related by analytic continuation to those obtained in the
$\epsilon \neq 0$ region where the sum is convergent.

The dangerous-looking term in  expression \fzlq\ for $ \epsilon
\to 0$ can be rewritten \eqn\dgr{ {\rm Im}[\cot ({n\pi \over
\beta^2}-n\pi i\epsilon)]={1\over 2} {\sinh (2n\pi \epsilon) \over
\sinh^2 (n\pi \epsilon)+ \sin^2 ({n\pi \over \beta^2})}.} To
define the limit we must take $\beta^2$ irrational so that $ \sin
({n\pi \over \beta^2}) $ is nonzero for every  $n$. In that case,
the $\sin^2 ({n \pi \over \beta})$ dominates over the $\sinh^2
(n\pi \epsilon)$ term for $\epsilon \to 0$. Because of the
 $\sinh (2n\pi \epsilon)$ in the numerator
every term in the sum vanishes for $\epsilon=0$ and hence
\eqn\dagr{ \lim_{\epsilon \to 0} {\rm Im}[\cot ({n\pi\over  \beta^2}-n\pi i\epsilon)]=0.}
 This leaves us with for $\epsilon \to 0$
\eqn\fstg{{\rm Re}[\Sigma_{i\beta}]=\sum_{n=1}^{\infty}{ e^{2\pi ny}
\over  2n} +\sum_{n=1}^{\infty}{ e^{-2\pi ny} \over  2n}.} Using
analytic continuation in $y$ to define the sums, and restoring
$x=\beta y$ gives \eqn\ftg{{\rm Re}[\Sigma_{i\beta}]=- \ln[ 2 \left(\sinh
{\pi x \over \beta}\right)].}

 Although derived for irrational $\beta^2$, this result
can obviously be smoothly extended back to the reals. For $\beta
\to 1$, the integral \ntr\ is real, and \ftg\ is the only real
part of $\ln S_{i}$. This then yields for $b\to i$\eqn\cpy{{\rm
Re}\ln [S_{i}(x)]=-\ln [2 \sinh (\pi x )].}

Now we consider the imaginary part of the sum. Here it is useful
to consider $\beta^2={q \over p}$ rational (with$(p,q)$ coprime).
The imaginary part is then \eqn\xlq{{\rm Im}  [
\Sigma_{i\beta,\epsilon}]=-\sum_{n=1}^{\infty}{ 1 \over n} \sinh
(2\pi n y ){\rm Re}[\cot (n\pi{p \over q}-n\pi i\epsilon)],} where
\eqn\dgr{ {\rm Re}[\cot (n\pi{p \over q}-n\pi i\epsilon)]={1\over
2} {\sin({2np\pi \over q}) \over \sinh^2 (n\pi \epsilon)+ \sin^2
({np\pi \over q})}.} Now we find that the terms with $n$ a
multiple of $q$ vanish, while the remaining terms are bounded but
typically nonvanishing for $\epsilon \to 0$. The sum is then for
$\epsilon =0$ \eqn\xlq{{\rm Im}  [\Sigma_{i\beta}]=-\sum_{n \neq
mq}^{\infty}{ 1 \over  n} \sinh (2\pi n y) \cot (n\pi{p \over
q}).} Again we will define the $e^{2\pi n y}$ terms by analytic
continuation from negative $y$ but for the sake of brevity we will
not bother to separate the two types of terms. Writing $n=mq+n_0$,
with $n_0=1,...(q-1)$, \xlq\ may be rewritten
\eqn\xclq{\eqalign{{\rm Im} [\Sigma_{i\beta}]&=-
\sum_{n_0=1}^{q-1}\sum_{m=0}^{\infty} {\sinh( 2\pi(mq+n_0){x \over
\beta})\over mq+n_0} \cot (n_0\pi{p \over q})\cr
 &=\int^{x \over \beta}_{-x \over \beta}  dy {\pi \over
e^{-2\pi q y}-1} \sum_{n_0=1}^{q-1} \cot (n_0\pi{p \over
q})e^{-2\pi n_0y} .}} The integral has an unilluminating
expression in terms of hypergeometric
 functions.
Note that for the case of current interest $b=i$, $p=q=1$ and
\xclq\ trivially vanishes. \xclq\ is a finite expression which
(together with \ntr) defines the phase of $S_b(x)$ for real $x$
and $b$ on the imaginary axis. It is a smooth function of $x$ for
any rational $b$. It is easy to see however\foot{For example
$q=p+1$ for large $p$ does not approach $q=p=1$.}, that it is not
a smooth function of $\beta^2$: it varies chaotically over the
rationals, and has no obvious extension to the reals.

Now let us consider the product $Y$ appearing in \prodx. This can
be written in terms of $S_b$ as
 \eqn\ywt{Y_b(\o)={S_b(2\o) \over
S_b^2(\o)}{b\over 2\pi} {\Gamma^2(1-{\o \over
b})\Gamma^2(-b\o)\over  \Gamma(1-{2\o \over b})\Gamma(-2b\o)}.} It
then follows from \ntr, \ftg\ and \xclq\ that
\eqn\fgi{Y_i(\o)=-e^{i\theta(\o)},} which is a pure phase in
agreement with the naive expectation from the product formula
\prodx. The phase is determined by \ntr\ as \eqn\rfv{ \theta=-i
I_i(2\o)+2iI_i(\o)= {1 \over 2}\int_0^{\infty} {d \tau\over
\tau}{\sin(4 \o \tau)-2\sin(2\o\tau)  \over \sinh^2(\tau) }.}
Hence, our continuation prescription yields \eqn\vvc{ d_i(\o) =
{(\pi \lambda)^{2i\omega}e^{i \theta(\o)}\over 4
  \cosh^2(\pi \o)} .}
This agrees asymptotically for large $\o$ with the minisuperspace
result for the string creation rate \dkaw.

It is interesting to consider the results of taking other contours
from ${\rm Re}b>0$ to $b=i$. Consider for example taking $b \to i$
along the arc $b=e^{i\theta}$ for $0\le \theta \le {\pi \over 2}$,
which has real $Q=2\cos \theta$. In this case the phase of $Y$ is
smooth (in fact it vanishes) but the magnitude fluctuates wildly
as a function of $\omega$ for $\theta \to {\pi \over 2}$. This can
be seen from the recursion relation\foot{Physically the recursion
formulae are derived by considering insertions of degenerate
operators.} \eqn\drft{\eqalign{Y_{b}(\o+Q)&= {4 (\o+Q)\over
\o}\bigl|{\Gamma(-2b\o)\Gamma^2(-b(\o+Q)) \over
\Gamma(-2b(\o+Q))\Gamma^2(-b\o)}\bigr|^2 \bigl| \cot(\pi b\o)\sin
\pi b (2\o+Q)\bigr|^2 Y_{b}(\o).}} For $b\to i$, $Q \to 0$, and
this reduces to \eqn\mml{Y_i(\o +0)=16 \cosh^4(\pi \o)Y_i(\o).}
Hence along this contour the two-point function does not smoothly
approach \vvc, in accord with the general expectation that
correlators at $b=i$ will depend on a contour prescription.

\centerline{\bf Acknowledgments} The work of A.S.  was supported
in part by DOE grant DE-FG02-91ER40654 The work of M.G.\ was supported by NSF
grant 9870115. We are grateful to I. Affleck, V. Kazakov, M. Kleban,
I. Kostov, D. Kutasov, J. Maldacena, A. Maloney, S. Minwalla,
D. Nelson, J. Polchinski, A. Sen, S.
Shenker, J. Teschner,
  X. Yin, A. Zamolodchikov and especially V. Schomerus and A.
Recknagel for useful discussions. M.G. gratefully acknowledges the
hospitality of LPTHE, Paris and especially B. Pioline during the
finalization of this paper.

\appendix{A}{Properties of special functions}
The special function $G_b$ was defined in \FateevIK\
(see also \PonsotNG). It obeys the recursion relations
\eqn\defbga{G_b(x+b)=(2\pi)^{-{1\over 2}} b^{-bx+{1\over 2}}
\Gamma(bx)G_b(x),\quad \quad
G_b(x+{1\over b})=(2\pi)^{-{1\over 2}}
b^{{x\over b}-{1\over 2}}\Gamma({x\over b})G_b(x). }
There exists an integral representation for $G_b(x)$, which is valid
for $Re(x)>0$,
\eqn\intresa{\ln G_b(x)= \int_0^\infty {dt\over t} \Big( {e^{-{Qt\over
2}}-e^{-xt}\over (1-e^{-bt})(1-e^{-{t\over b}})} +{({Q\over 2} -x)^2
\over 2}e^{-t} +{{Q\over 2}-x\over t}\Big).}
Another useful special function is defined by  $S_b(x)= G_b(Q-x)/G_b(x)$
and  satisfies the recursion relations
\eqn\defbgb{\eqalign{S_b(x+b)&= 2 \sin(\pi b x)S_b(x),\cr   S_b(x+{1\over
b} )&= 2 \sin(\pi  {x\over b})S_b(x),\cr S_b(x+Q)&=-4
 \sin(\pi b x) \sin(\pi  {x\over b})S_b(x),  }}
as well as
\eqn\mvg{S_b(x)S_b(-x)=-{1 \over 4 \sin (\pi b x )\sin(\pi  {x\over b})}.}
Since $Q=b+{1\over b}$ one can easily show by applying \defbga\ twice
that
\eqn\defgbc{\eqalign{G_b(x+Q)&= {b^{{x\over b}- bx+1}\over 2\pi} \Gamma({x\over
    b}+1)\Gamma( b x) G_b(x),\cr
G_b(x-Q)&= {b^{-{x\over b}+ bx+{1\over b^2}-b^2 -1 }2\pi \over  \Gamma({x\over
    b}-{1\over b^2})\Gamma( b x-1-b^2)} G_b(x). }}

$G_b(x)$ is related to a the
Barnes double Gamma function $\Gamma_2(x|\nu_1,\nu_2)$  \Barnes\ in
the following way

\eqn\barnrel{G_b(z)=\Gamma_2^{-1}(z|b,1/b). }
 The Barnes
double Gamma function is related to the double Hurwitz function
\refs{\Barnes,\JimboSS}
\eqn\hurw{\zeta_2^s(x|\nu_1,\nu_2)= \sum_{m,n>0} (n\nu_1+m\nu_2+x)^{-s},}
in the following way
\eqn\barnhur{\Gamma_2(x|\nu_1,\nu_2) = \exp \big( {\partial\over
\partial s} \zeta_2^s(x|\nu_1,\nu_2) |_{s=0}\big). }
It follows from \hurw\ and  \barnhur\ that $\Gamma_2$
has a product representation
\eqn\proda{
\Gamma_2^{-1}(z|\nu_1,\nu_2)= e^{{z^2\over 2 \gamma_{21}} +
z\gamma_{22}}z \prod_{m=0}^\infty \prod_{n=0}^{\infty }\big(1+
{z\over
  \Omega}\big)e^{-{z\over \Omega}+ {z^2\over 2\Omega^2}},}
where $\Omega= m\nu_1+n\nu_2$ and $\gamma_{21},\gamma_{11}$
are functions of $\nu_1,\nu_2$ but not $z$ which can be found in
\Barnes.
It follows from \proda\ that
\eqn\prodz{{G_b(-z)\over G_b(z)}= -  e^{ +2
z\gamma_{22}} \prod_{m=0}^\infty \prod_{n=0}^{\infty }\big({z+
  \Omega\over -z+\Omega }\big)e^{-{2z\over \Omega} }.}
In the limit $x\to \infty$ with $\pm Im(x)>0$ one finds \refs{\Barnes,
\JimboSS},
\eqn\limsb{\ln \big( S_b(x)\big)= \pm i\pi \Big( {x^2\over 2} - {Q
x\over 2}-{1\over 2}(b^2+{1\over b^2}+2)\Big)+o({1 \over x}).}

\listrefs

\end